\documentclass[12pt,twoside,a4paper]{article}
\usepackage{amsmath,amssymb,latexsym,theorem,natbib,epsfig,color,subfigure}
\usepackage{multirow,graphicx,array,rotating,multicol}  
\usepackage{epstopdf,url,afterpage, bbm}

\def\crps{\mathop{\hbox{\rm CRPS}}}

\numberwithin{equation}{section}

\title{Statistical post-processing of ensemble forecasts of temperature
  in Santiago de Chile}

\author{{\sc Mailiu D\'{\i}az$^{a}$}, {\sc Orietta Nicolis$^{a,b}$}, {\sc Julio C\'esar Mar\'\i n$^{c}$} \\ and {\sc S\'andor Baran$^{d}$} \vspace*{0.5cm}\\
  $^a$Department of Statistics, University of Valpara\'\i so \\ Gran Breta\~na 1111, Valpara\'\i so, Chile \\
  $^b$Faculty of Engineering, Andres Bello University\\Quillota 980, Vi\~na del Mar, Chile\\
    $^c$Department of Meteorology, University of Valpara\'\i so \\ Gran Breta\~na 644, Valpara\'\i so, Chile\\
         $^d$Faculty of Informatics, University of Debrecen\\
         Kassai \'ut 26, H-4028 Debrecen, Hungary 
        }

\date{}
        
\begin{document}
\pagestyle{myheadings}

\maketitle

\begin{abstract}
Currently all major meteorological centres generate ensemble forecasts using their operational ensemble prediction systems; however, it is a general problem that the spread of the ensemble is too small, resulting in underdispersive forecasts, leading to a lack of calibration. In order to correct this problem, different statistical calibration techniques have been developed in the last two decades.

In the present work different post-processing techniques are tested for calibrating 9 member ensemble forecast of temperature for Santiago de Chile, obtained by the  Weather Research and Forecasting (WRF) model using different planetary boundary layer and land surface model parametrization. In particular, the ensemble model output statistics (EMOS) and Bayesian model averaging techniques are implemented and since the observations are characterized by large altitude differences, the estimation of model parameters is adapted to the actual conditions at hand.

Compared to the raw ensemble all tested post-processing approaches significantly improve the calibration of probabilistic and the accuracy of point forecasts. The EMOS method using parameter estimation based on expert clustering of stations (according to their altitudes) shows the best forecast skill.

\end{abstract}

\section{Introduction}
\label{sec:sec1}
The central zone of Chile, located between 32$^{\circ}$ and 37$^{\circ}$ south latitude, has a semi-arid Mediterranean climate as a result of the influence of topographic barriers and the South East Pacific Anticyclone located over the cool Pacific Ocean, which generates a persistent inversion layer in the lowest few hundred meters of the atmosphere \citep{Burger2018}. In particular, Santiago the Chile surrounded by high mountains has its own special micro climate making accurate weather forecasts even more complicated. 

Obtaining reliable forecasts of surface temperature has a large impact in many fields such as renewable energies, air quality and radiative transfer. These forecasts can be obtained with the help of numerical weather prediction (NWP) models which provide predictions on high spatial and temporal resolutions. In the last few years in Chile serious efforts have been made in evaluating meteorological models to improve temperature forecasts, among other variables  \citep{Saide2011,Cortes2011,Pozo2016,Gonzalez2017}. However, the outputs of these NWP models are subject to an intrinsic uncertainty, which can be controlled and reduced by running the models with different initial conditions and parametrizations opening the door for ensemble forecasting \citep{leith}.
Currently all major meteorological centres generate ensemble forecasts using their operational ensemble prediction systems (EPSs). Examples of this are the 51 member EPS of the European Centre for Medium-Range Weather Forecasts \citep{ecmwfens,lp08} and the 30 member Consortium for Small-scale Modelling EPS
of the German Meteorological Service \citep{gtpb}. A general problem with many operational EPSs is that the spread of the ensemble is too small, resulting in underdispersive forecasts, leading to a lack of calibration \citep[see e.g.][]{bhtp,ppb}.

A possible way of improving ensemble forecasts is the use of some form of statistical post-processing. In the last 15 years several different statistical calibration techniques have been developed \citep[for an overview see][]{wfk,ppb} including non-homogeneous regression or ensemble model output statistics \citep[EMOS;][]{grwg} and Bayesian model averaging \citep[BMA;][]{rgbp}, both providing full predictive distribution of the future weather quantity. The EMOS predictive distribution is specified by a parametric distribution family with parameters connected to the ensemble members via appropriate link functions, whereas BMA applies mixture distributions with components corresponding to the ensemble members. Given a predictive distribution, as point forecasts, one can consider either its mean or its median, and probabilities of various events can also be easily calculated. EMOS and BMA models corresponding to various weather quantities differ in the parametric laws they are based on. A normal distribution or normal mixture fits well temperature and pressure forecasts \citep{grwg,rgbp}, wind speed requires a non-negative and skewed distribution such as truncated normal \citep{tg,bar}, log-normal \citep{bl15} or gamma  \citep{sgr10}, whereas to calibrate precipitation accumulation, a discrete-continuous model with point mass at zero is required \citep{srgf,sch,schham,bn}.

In the present work various post-processing models for calibrating ensemble forecasts of temperature in Santiago city are evaluated. The ensemble members correspond to nine Weather Research and Forecasting \citep[WRF;][]{wrfmod} model configurations with three nested domains. According to our best knowledge, no studies have been published yet with the aim of improving the quality of surface temperature predictions in Chile by statistical calibration based on ensemble post-processing techniques. Moreover, as the results show, the location of the ensemble domain with large altitude differences requires some adaptation of the post-processing approach to the actual conditions at hand.

The paper is organized as follows. A description of WRF configurations, data from meteorological stations, and their preliminary statistical analysis is provided in Section \ref{sec2}. Section \ref{sec3} describes the post-processing models and applied methods of model verification, whereas the results of statistical post-processing are given in Section \ref{sec4}.   Finally, Section \ref{sec5} concludes the paper with a summary of the major findings and a discussion of possible future areas of research.

\section{WRF configurations and data description}
\label{sec2}

The WRF model is employed to generate nine different simulations resulting in a 9 member forecast ensemble for surface temperature (K) with a lead time of 24 h for the period between 1 October 2017 and 30 January 2018 every 3 hours from 00 UTC to 21 UTC. The corresponding verifying observations are obtained from 19 meteorological stations around Santiago city.   

\subsection{WRF simulations}
 \label{subs2.1}
\begin{figure}[t]
	\centerline{ \begin{tabular}{cc}
			\includegraphics[height=160pt]{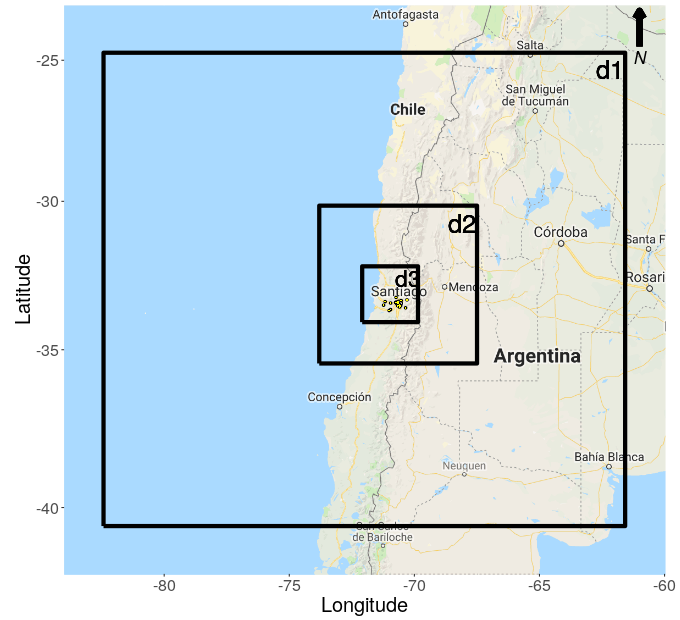}& \includegraphics[height=170pt]{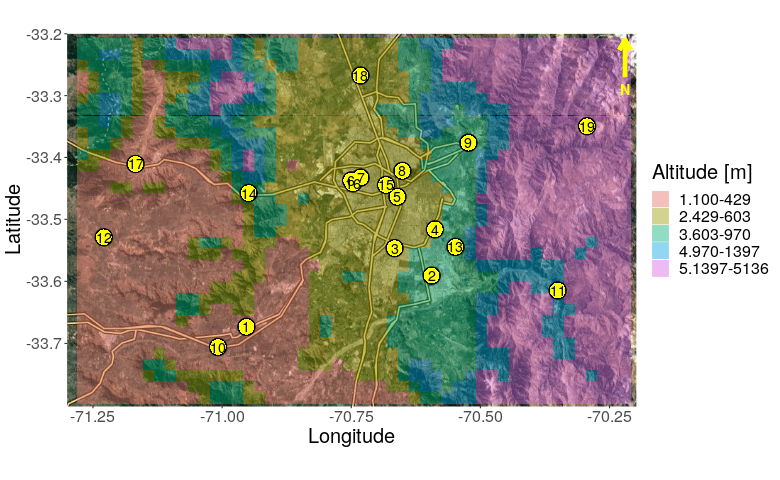} \\
			{\small (a)} &{\small (b)} 
	\end{tabular}}
	\caption{Location maps: (a) WRF domains 1, 2, and 3; (b) domain 3 with terrain elevation (m) and location of 19 meteorological stations.}
	\label{map}
\end{figure}

Three model nested domains (see Figure \ref{map}a), at 18 km, 6 km and 2 km horizontal resolutions, were employed in the simulations using Version 3.7.1 of the Advanced Research WRF core \citep[ARW-WRF;][]{Skamarock2008}. Results from the highest resolution domain (d3) were used in this study with a superficial area of 208$\times$208 km. 

The simulations included 44 vertical levels with variable resolution, with eight levels within the first kilometer in the vertical with spacing varying between 60 and 200 m. The Final Analysis (FNL) at 0.25 deg. $\times$ 0.25 deg. horizontal resolution provided the initial and boundary conditions every 6 hours, and an analysis nudging was implemented in the outer domain (d1) to provide better boundary conditions during the simulation period. Table \ref{schemes} describes the parametrizations used on each simulation. 

\begin{table}[t]
	\caption{Description of the parametrizations set used on each of the nine members.}
	\begin{center}
		\begin{tabular}{cccccccc}
			\hline
			Member	&
			LSM&Surface&PBL&
			Cumulus&Microph.&
			LW Rad&	SW rad\\
			\hline
			1&Noah &MYNN&
			MYNN&Kain-F &
			WSM3 &RRTMG &
			RRTMG \\
			2&Noah &	MYJ &
			MYJ &Kain-F &
			WSM3 &RRTMG &
			RRTMG \\
			3&Noah &YSU&
			YSU&Kain-F &
			WSM3 &RRTMG &
			RRTMG \\
			4&Pleim-Xiu&MYJ&
			MYJ&Kain-F &
			WSM3 &RRTMG &
			RRTMG \\
			5&Pleim-Xiu&
			YSU&YSU&
			Kain-F &WSM3 &
			RRTMG &RRTMG \\
			6&5-layer&MYJ&
			MYJ&Kain-F &
			WSM3 &RRTMG &
			RRTMG \\
			7&5-layer&YSU&
			YSU &Kain-F &
			WSM3 &RRTMG &
			RRTMG \\
			8&5-layer&MYNN &	MYNN &Kain-F&
			WSM3 &RRTMG &
			RRTMG\\
			9&Pleim-Xiu&MYNN&
			MYNN&Kain-F &
			WSM3 &RRTMG&
			RRTMG\\
			\hline
			\multicolumn{8}{l}{LSM: Land-surface model, PBL: planetary boundary layer scheme,}\\
			\multicolumn{8}{l}{Microph: microphysics scheme, Cumulus: represents the convective scheme,}\\
			\multicolumn{8}{l}{LW Rad and SW Rad represent the longwave and shortwave radiation schemes.}
		\end{tabular}
	\end{center}
	\label{schemes}
\end{table}

The members differ each other in the applied planetary boundary layer (PBL) and land-surface model (LSM) parametrizations. We use the Mellor-Yamada-Janjic \citep[MYJ;][]{Janjic1994}, Yonsei University \citep[YSU;][]{Hong2006} and Mellor-Yamada Nakanishi and Ninno 2.5 level \citep[MYNN;][]{Nakanishi2006} schemes to represent the PBL and surface layer processes. The Land-surface processes are represented by the 5-layer \citep{Dudhia1996}, Noah \citep{Chen2001} and Pleim-Xiu \citep{Pleim2003} schemes. The rest of the parametrizations are kept the same in all simulations. We use the Kain-Fritsch \citep[Kain-F;][]{Kain2004} cumulus parametrization, the Rapid Radiative Transfer Model \citep[RRTMG;][]{Iacono2008} to represent the longwave and shortwave radiative processes and the WRF single-moment 3-class\citep[WSM3;][]{Hong2004} scheme to represent convective processes.

\subsection{Station observations}
\label{subs2.2}

Surface temperature observations (K) every 3 hours for the period between 1 October 2017 and 30 January 2018 were obtained from the Direcci\'{o}n Meteorol\'ogica de Chile (DMC, Government agency responsible for managing the meteorological data  and providing operational weather forecasts in the country, available online at \url{http://www.meteochile.gob}), and from the National System for Air Quality (Environmental Ministry, see \url{https://sinca.mma.gob.cl/}) for 19 meteorological stations described in Table \ref{stations} (see also Figure \ref{map}b).

\begin{table}[t]
	\caption{Geographical coordinates in decimal degrees and altitude of monitoring stations.}
	\label{stations}
\begin{center}
	\begin{tabular}{llccc}
		\hline
		No. &Station &Longitude  &Latitude    &Altitude (m)\\ 
		\hline
		1&Talagante&-70.9528&-33.6733&390\\
		2&Puente Alto&-70.5944&-33.5908&670\\
		3&El Bosque&-70.6660&-33.5465& 580\\
		4&La Florida&-70.5881&-33.5161&601\\
		5&Parque O'Higgins&-70.6606&-33.4637&549\\
		6&Pudahuel&-70.7501&-33.4373&494\\
		7&Cerro Navia&-70.7319&-33.4326&500\\
		8&Independencia&-70.6511&-33.4217&560\\
		9&Las Condes&-70.5232&-33.3763&798\\
		10&El Paico&-71.0079&-33.7059&275\\
		11&San Jos\'{e} Guayac\'{a}n&-70.3505&-33.6148&928\\
		12&Chorombo Hacienda&-71.2291&-33.5289&145\\
		13&Aguas Andinas&-70.5483&-33.5445& 665\\
		14&Lo Prado &-70.9485&-33.4575&1068\\
		15&Quinta Normal&-70.6827&-33.4445&534\\
		16&San Pablo&-70.7466&-33.4417&490\\
		17&Curacav\'i&-71.1676&-33.4106&208\\
		18&Lo Pinto&-70.7321&-33.2675&512\\
		19&El Colorado&-70.2935&-33.3495&2750 \\
		\hline
        \end{tabular}
      \end{center}
\end{table}

We remark that Las Condes (9), La Florida (4) and Chorombo Hacienda (12) stations have more than 25\,\% of missing values.     

\subsection{Predictive performance of the raw ensemble}

\begin{figure}[ht]
	\centerline{\includegraphics[width = .8\textwidth]{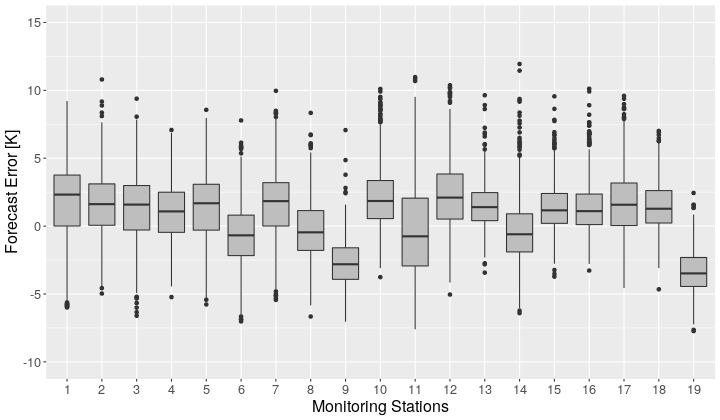}} 
	\caption{Forecast error of the ensemble mean for different monitoring stations.}
	\label{errStat}
\end{figure}

\begin{figure}[t]
  \centerline{\includegraphics[width = .49\textwidth]{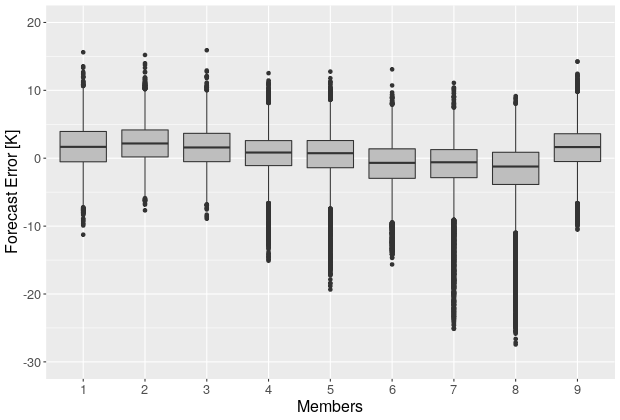} \\
    \includegraphics[width = .49\textwidth]{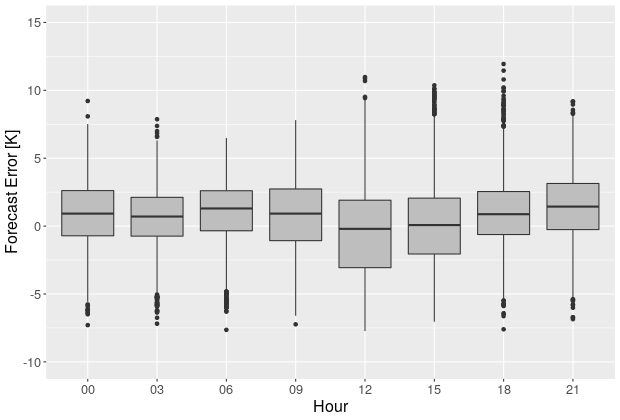}}
  \centerline{\hbox to 9 cm{\small (a) \hfill (b)}}
	\caption{Forecast error (a) of the different ensemble members for all dates and forecast hours; (b) of the ensemble mean for different forecast hours for all dates.}
	\label{errMemHour}
\end{figure}

As the altitude description of stations in Table \ref{stations} and Figure \ref{map}b show, the topography within the ensemble domain of study is very complex causing difficulties to obtain reliable forecasts. Figure \ref{errStat} provides box-plots of the forecast errors of the ensemble means for the different stations. It shows that the raw ensemble systematically underestimates temperature at stations located at higher altitudes (stations 9, 11, 14 and 19). This fact should be obviously taken into account during the calibration process. A number of studies have found cold biases in near-surface temperature forecasts with the WRF model in different mountainous regions in Chile and other parts of the world using similar options for PBL and LSM schemes as those used in this study \citep{rsnp10, mpmtc13,mskc16, Gonzalez2017}. The temperature underestimation may be the result of misrepresentations in the real orography or the near-surface moisture in the model \citep{mskc16}. There is also some variation in the performance of the different ensemble members (Figure \ref{errMemHour}a) and the forecast hour has influence on the forecast error (Figure \ref{errMemHour}b), too. According to Figure \ref{errMemHour}a ensemble members 7 and 8 underperform the other 7 forecasts, whereas the performance of forecasts for 12 and 15 UTC also differs from the results for other forecast hours. The latter might be related to the misrepresentation of the real orography in the model. The temperature diurnal variation is strongly influenced by solar radiation, which strongly varies in sites located in complex terrain due to topographic shading and variations in slope orientations \citep{zhang18}. Therefore, the time of sunrise in those stations might be misrepresented, causing larger errors at above mentioned hours.

\begin{figure}[t]
	\centerline{\includegraphics[width = .45\textwidth]{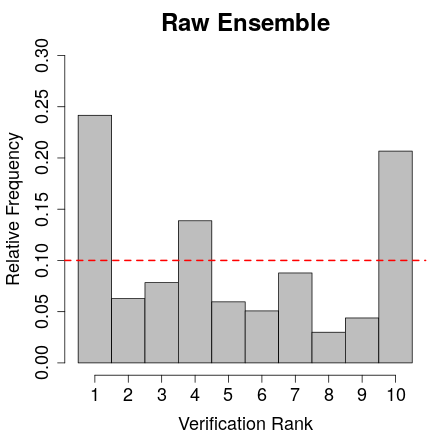}} 
	\caption{Verification rank histogram of raw ensemble forecasts for the period 1 October 2017 -- 30 January 2018.}
	\label{verRankAll}
\end{figure}

In order to get a first overview of the raw ensemble calibration, one can have a look at the verification rank histogram (Figure \ref{verRankAll}), which represents the histogram of ranks of validating observations with respect to the corresponding ensemble forecasts computed from the ranks at all locations and dates considered \citep[see e.g.][Section 7.7.2]{wilks}. In case of a proper calibration the ranks should be uniformly distributed, which is far not the case in Figure \ref{verRankAll}, calling for some form of statistical post-processing.

\section{Models}
\label{sec3}

As mentioned in the Introduction, normal distribution is suitable for temperature modelling, hence we make use of normal EMOS and BMA models suggested by \citet{grwg} and \citet{rgbp}, respectively. In what follows, let $f_1, \ldots ,f_9$ denote the 9 ensemble members for a given location and time point.

\subsection{EMOS model for temperature}
 \label{subs3.1}
The normal EMOS predictive distribution in our case has the form
\begin{equation}
\label{eq:EMOS}
  \mathcal{N}(a_{0}+a_{1}f_{1}+ \cdots+a_{9} f_{9}, b_{0}+b_{1}S^2),\qquad \textrm{where} \qquad S^2:=\frac{1}{8}\sum_{k=1}^{9}(f_{k}-\bar{f})^2,
\end{equation}
with $\bar{f}$ denoting the ensemble mean. Location parameters $a_{0},a_{1},\ldots,a_{9}\in \mathbb{R}$ and scale parameters $b_{0},b_{1} \in \mathbb{R}$ ($a_1, \ldots ,a_9,b_0,b_1\geq 0$) are estimated by optimizing the value of a proper scoring rule (see Section \ref{sub3.3}), over the training data consisting of ensemble members and verifying observations from the preceding $n$ days. 

\subsection{BMA model for temperature}
\label{subs3.2}
EMOS calibration is efficient in the case of unimodal distributions, however, it cannot provide an appropriate model for weather variables following distributions with several modes. In such situations BMA modelling using mixture distributions might be of more use. The normal BMA model of \citet{rgbp} for calibrating temperature forecasts in our case results in the predictive PDF
\begin{equation}
  \label{eq:BMA}
p(x|f_{1},\ldots,f_{9}):=\sum_{k=1}^{9} \omega_{k} \frac 1\sigma \varphi \Big (\frac{x- \beta_{0,k} - \beta_{1,k}f_k}\sigma \Big),  
\end{equation}
where $\varphi$ denotes the PDF of the standard normal distribution. Here the weights $\omega_{k}$ reflect to the relative performance of each ensemble member $f_{k}$ during the training period and note that they should form a probability distribution, that is $ \sum_{k=1}^{9} \omega_{k} = 1$, $\omega_{k} \geq 0$. The linear form $\beta_{0,k} + \beta_{1,k}f_k$ of the mean of each component PDF is responsible for the bias correction, however, one might also consider cases $\beta_{1,k}=1$ (additive bias correction) and $\beta_{0,k}=0, \ \beta_{1,k}=1$ (no bias correction). The latter two approaches might be helpful in situations where the raw ensemble is unbiased and additional bias correction might introduce unnecessary extra errors \citep[see e.g.][]{bhn14}

Similar to the EMOS approach, location parameters $\beta_{0,k},\beta_{1,k}$, weights $\omega_k, \ k=1,\ldots,9$, and scale $\sigma$ are estimated using appropriate training data. Location parameters are obtained from regressing the validating observations on the ensemble members, whereas weights and scale are estimated using a maximum likelihood approach where the likelihood function is maximized with the help of the EM algorithm for mixtures \citep[see e.g.][]{mclk}. Note that BMA modelling is performed with the help of the {\tt ensembleBMA R}  package \citep{frgsb}.

\subsection{Verification scores}
\label{sub3.3}

The aim of statistical post-processing is to maximize the sharpness of the predictive distribution subject to calibration \citep{gbr}, where the former refers to the concentration of the forecast distribution, whereas the latter to the consistency between predicted probabilities and observed relative frequencies. These goals can be addressed simultaneously with the help of scoring rules \citep{grjasa} assigning numerical values to pairs of forecast distributions and validating observations. One of the most widely used proper scoring rule in atmospheric sciences is the continuous ranked probability score \citep[CRPS;][]{grjasa}. For a predictive cumulative distribution function (CDF) $F(y)$ and observation $x$  the CRPS is defined as
\begin{equation}
  \label{eq:CRPS}
\crps\big(F,x\big):=\int_{-\infty}^{\infty}\big (F(y)-{\mathbbm 
  1}_{\{y \geq x\}}\big )^2{\mathrm d}y={\mathsf E}|X-x|-\frac 12
{\mathsf E}|X-X'|, 
\end{equation}
where ${\mathbbm 1}_H$  denotes the indicator of a set $H$, while  $X$  and  $X'$ are independent random variables with CDF $F$  and finite first moment. Note that the CRPS can be expressed in the same units as the observation and it is a negatively oriented scoring rule, that is the smaller the better.

Further, calibration of a predictive distribution can be investigated using the coverage of the $(1-\alpha )100\,\%, \ \alpha \in (0,1),$ central prediction interval. By coverage we mean the proportion of validating observations located between the lower and upper  $\alpha /2$  quantiles of the predictive CDF and level $\alpha$ should be chosen to match the nominal coverage of the raw ensemble, which is $80\,\%$ for the 9 member ensemble at hand.  As the coverage of a calibrated predictive distribution should be around $(1-\alpha )100\,\%$, the suggested choice of $\alpha$ allows direct comparisons with the raw ensemble.

The improvement in calibration with respect to the raw ensemble can also be demonstrated with the help of the probability integral transform (PIT) histograms \citep{wilks}. The PIT is the value of the predictive CDF evaluated at the validating observation and in case of proper calibration it should follow a uniform law on $[0,1]$ interval. Hence the PIT histogram is the continuous counterpart of the verification rank histogram of the raw ensemble.

Finally, a predictive performance of point forecasts is evaluated with the help of mean absolute error (MAE) and root mean squared error (RMSE). Note, that the former is optimal for the median, whereas the latter is for the mean \citep{gneiting11,pinhag}. 

\subsection{Training data}
\label{subs3.4}

The choice of appropriate training data is essential for estimating the  parameters of the EMOS and BMA models given by \eqref{eq:EMOS} and \eqref{eq:BMA}. In general, rolling training periods are applied; however, there are two main approaches to choosing forecast cases for training \citep{tg}. In the local approach parameters for a given station are estimated from the data of that particular station only. This approach usually results in a very good model fit, provided the training period is long enough to avoid numerical issues in parameter estimation \citep[see e.g.][]{hemri14}. In contrast, regional EMOS and BMA estimate parameters using all available forecast cases from the training period, thus all stations in the forecast domain share the same set of parameters. In this way one can use shorter training periods, but the regional approach is not suitable for large heterogeneous domains, e.g. for global forecasts. Recently \citet{lb17} proposed a third, semi-local approach, which combines the advantages of local and regional forecasting. Training data for a given station are augmented with data for stations with similar characteristics, e.g. by clustering the stations using feature vectors determined by station climatology and/or ensemble forecast errors during the training period. Within a given cluster a regional parameter estimation is performed, but note that clusters may vary as the training window slides \citep[for more details see][]{lb17}. Finally, one can also perform an expert clustering of the monitoring stations based on their location or other covariates, and estimate the parameters within clusters regionally.

\section{Results}
\label{sec4}

\begin{figure}[t]
	\centerline{\includegraphics[width = .7\textwidth]{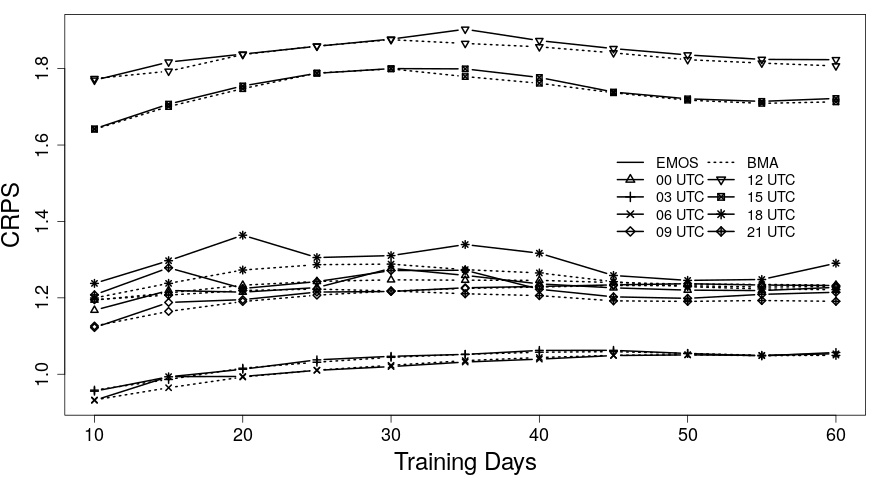}} 
	\caption{Mean CRPS of EMOS and BMA predictive distributions for the period 30 November 2017 -- 30 January 2018.}
	\label{crpsEMOS}
\end{figure}

In order to exclude the natural daily variation in temperature, EMOS and BMA calibration of WRF ensemble forecasts described in Section \ref{subs2.1} is performed separately for each forecast hour. Since the ensemble has 9 members, EMOS post-processing requires a total of 11 parameters to be estimated, whereas for BMA the number of free parameters is 27. The data set at hand covers only 122 calendar days which for both approaches makes the local estimation of parameters impossible. Regional estimation also requires at least a 6 day training period (114 forecast cases) for EMOS and at least a 15 day training (285 forecast cases) for BMA in order to have at least 10 forecast cases for each parameter.

\begin{table}[t]
        \caption{CRPS of probabilistic, RMSE of mean and MAE of median forecasts, and coverage of 80\,\% central prediction intervals.}
        \begin{center}
                \begin{tabular}{llccccccccc}
                        \hline
                        \multirow{2}{*}{Scores}&\multirow{2}{*}{Models}&
                        \multicolumn{8}{c}{Forecast hour (UTC)}&\multirow{2}{*}{Overall}\\
                        &&00 &03 &06&09&12&15&18&21&\\
                        \hline
&Ens&1.728& 1.570&1.743&1.773&2.135&1.839&1.604&1.764&1.769 \\                  CRPS&EMOS&1.186&1.012&1.017&1.219&1.775&1.672&1.309&1.260&1.306\\
(K)&EMOS-C&1.105&0.955&0.992&1.218&1.686&1.505&1.163&1.180&1.225\\
&BMA&1.197& 1.018& 1.019& 1.221& 1.778& 1.668& 1.247& 1.254&1.299\\ \hline                     
&Ens&2.794&2.569&2.817&2.835&3.265&3.297&3.041&3.019&2.963\\
RMSE&EMOS&2.092&1.791&1.808&2.156&3.121&2.988&2.418&2.228&2.370\\
(K)&EMOS-C&1.951&1.692&1.755&2.149&2.996&2.697&2.111&2.114&2.222\\
&BMA&2.097& 1.795& 1.810& 2.156& 3.118& 2.959& 2.236& 2.212&2.342\\ \hline                    
&Ens&2.255&1.903&2.153&2.312&2.712&2.496&2.244&2.500&2.321\\                 MAE&EMOS&1.683&1.417&1.437&1.729&2.557&2.367&1.828&1.788&1.850\\                (K)&EMOS-C&1.571&1.340&1.393&1.727&2.415&2.104&1.592&1.659&1.725\\
&BMA&1.693& 1.420& 1.443& 1.732& 2.533& 2.357& 1.745& 1.778& 1.837 \\ \hline
&Ens&51.15& 58.14& 57.05& 61.14&40.46&69.04&80.98&67.18&60.63 \\
Cover.&EMOS&75.63&75.56&76.36&76.66&77.30&77.59&75.37&78.36&76.60\\
(\%)&EMOS-C&76.40&74.25& 75.53&75.29& 76.86& 73.84& 72.62& 76.88&75.21\\
&BMA&74.42& 76.33& 77.29&75.12& 74.49& 75.61& 76.09& 76.44&75.72  \\
                        \hline
                \end{tabular}
        \end{center}
        \label{scores}
      \end{table}

We determine the length of the optimal training period by calibrating the ensemble forecasts using training periods of length $10,15,\ldots ,60$ days and comparing the predictive performance on the verification period 30 November 2017 -- 30 January 2018 (62 calendar days). Figure \ref{crpsEMOS} shows the mean CRPS values of EMOS and BMA predictive distributions as functions of the training period length for all forecast hours. Both  models have the best predictive performance at forecast hours 3 and 6 UTC, provide a bit higher CRPS values for 0, 9, 18 and 21 UTC, whereas the worse forecast skill corresponds to 12 and 15 UTC. However, these results are partially in line with Figure \ref{errMemHour}b and might be explained by the differences in accuracy of ensemble forecasts for different periods of a day. Note that the curves have their minima at day 10 and except the ones corresponding to 12 and 15 UTC do not show much variability. MAE and RMSE values of EMOS and BMA median and mean forecasts (not reported), respectively, are very consistent to the CRPS and do not change the overall picture. As in general one prefers shorter training periods, but has to keep in mind also the minimum number of forecast cases for parameter estimation, a training period of length 15 or 20 days should be selected. Finally, a 20 day training period is chosen for calibrating the WRF forecasts as it might also be appropriate for semi-local estimation of EMOS parameters using 2 or 3 clusters of observation stations. In this way the predictive performance of EMOS and BMA post-processed forecasts can be tested on temperature data from the period between 21 October 2017 and 30 January 2018 (102 calendar days).

Besides regional estimation of BMA and EMOS parameters, clustering based semi-local EMOS parameter estimation is also tested with 2 and 3 clusters and 24 features. Half of the features are obtained as equidistant quantiles of the climatological CDF, whereas the other half as equidistant quantiles of the empirical CDF of the forecast error of the ensemble mean over the training period \citep{lb17}. However, this approach often produces very unbalanced cluster sizes even for two clusters. In almost 28\,\% of the cases, station 19 alone forms a separate cluster, whereas the other cluster consists of the remaining stations. In particular, for 18 UTC this was the case on 76 out of 102 days in the training period. This uneven clustering is in line with the bad ensemble forecast skill at station 19 (Figure \ref{errStat}). Having a single station in a cluster means local parameter estimation for that particular station resulting in numerical issues in the optimization procedure.

\begin{figure}[t]
	\centerline{\includegraphics[width = .65\textwidth]{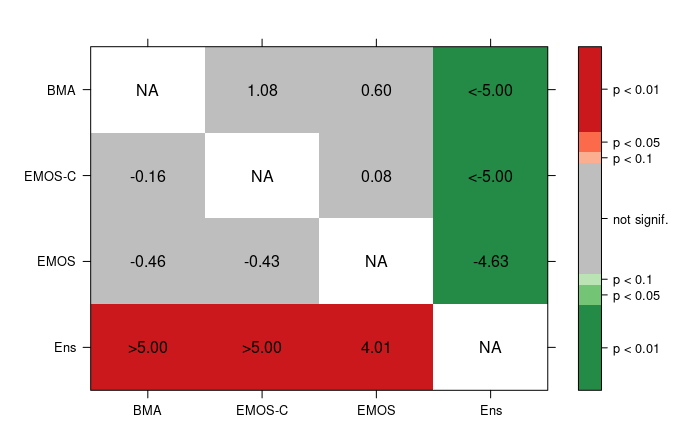}} 
	\caption{Values of the test statistics of the DM test for equal predictive performance based on the CRPS ({\em upper triangle}) and AE of the median forecast ({\em lower triangle}). Negative/positive values indicate a superior predictive performance of the forecast given in the row/column label, whereas green/red background indicates significant differences.}
	\label{dmPlot}
\end{figure}

Instead of grouping the stations dynamically based on feature vectors, one might try some form of expert clustering and for the ensemble domain at hand (see Figure \ref{map}b), altitude might be a reasonable covariate. The chosen altitude regions resulting in 3 clusters are: under 400 m (stations 1, 10, 12, 17); between 400 m and 750 m (stations 2, 3, 4, 5, 6, 7, 8, 13, 15, 16, 18);
above 750 m (stations 9, 11, 14, 19). Note that at stations in the third cluster, WRF forecasts systematically underestimate temperature (Figure \ref{errStat}).

\begin{figure}[t]
  \centerline{\hbox{\includegraphics[width = .4\textwidth]{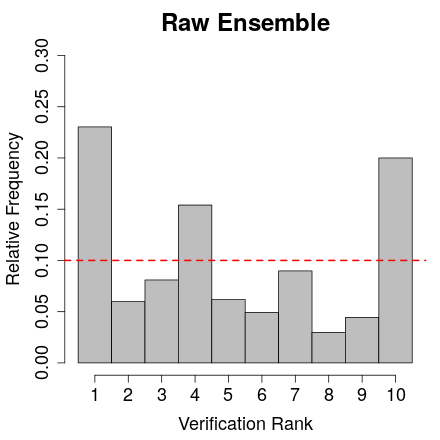}}
    \qquad \includegraphics[width = .4\textwidth]{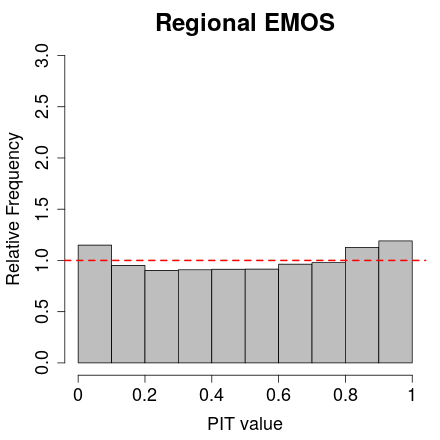}}
   \centerline{\hbox{\includegraphics[width = .4\textwidth]{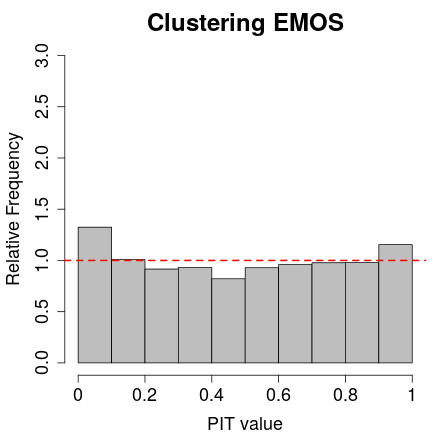}}
  \qquad \includegraphics[width = .4\textwidth]{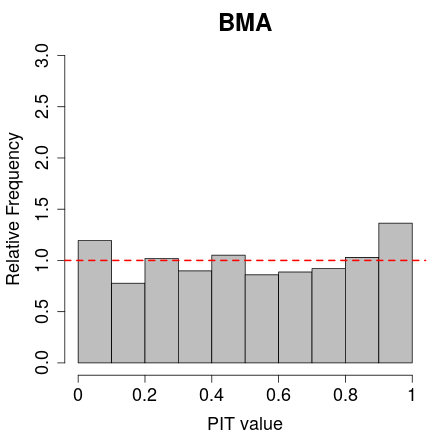}}
	\caption{Verification rank histogram of the raw ensemble and PIT histograms of post-processed forecasts for the period 21 October 2017 -- 30 January 2018.}
	\label{pit}
\end{figure}

In this way for post-processing ensemble forecasts we consider BMA and EMOS models with regional parameter estimation and EMOS model with clustering (EMOS-C) using a 20 day rolling training period. Table \ref{scores} shows the mean CRPS of probabilistic, the RMSE of mean and the MAE of median forecasts and the coverage of 80\,\% central prediction intervals separately for the different forecast hours and for the overall verification period. Note that all post-processing approaches outperform the raw ensemble in terms of all scores for all forecast hours but the coverage at 18 UTC; however, even for that forecast hour, the values are not far from the nominal 80\,\%. From the three competing post-processing approaches clustering based EMOS shows the best predictive performance followed by the BMA and regional EMOS. However, the differences between the scores of the three post-processing approaches are minor.

The statistical significance of differences between overall mean CRPS and MAE values is investigated with the help of two-sided  Diebold-Mariano \citep[DM;][]{dm95} tests of equal predictive performance, as this test takes into account temporal dependence \citep[for details of application see e.g.][]{bl18}. Figure \ref{dmPlot} shows the values of the DM test statistics based on the CRPS and the absolute error (AE) of median for all pairwise comparisons of forecasts. All calibration methods result in significant improvement compared to the raw ensemble; however, the forecast skill of the three post-processing approaches does not differ significantly.
This means that for the WRF forecasts at hand, the use of a more complex model for calibration with many parameters does not necessary pay off. One should use a more simple approach instead and possibly group the stations in a careful way.

\begin{table}[t]
	\caption{p-values of Kolmogorov–Smirnov tests for uniformity of PIT values. Average of 1000 random samples of sizes 1000 each.}
	\begin{center}
          \begin{tabular}{lccc}\hline
            Model&EMOS&EMOS-C&BMA \\ \hline
            Mean $p$-value&0.119& 0.104& 0.058\\ \hline
          \end{tabular}
        \end{center}
        \label{KStab}
      \end{table}
      
The same conclusions can be drawn from the overall PIT histograms plotted in Figure \ref{pit}. The improvement in calibration compared to the raw ensemble is obvious; however, all three PIT histograms are slightly U-shaped indicating a very small underdispersion (see also the overall coverage values of Table \ref{scores}). Unfortunately, the Kolmogorov–Smirnov test rejects the
uniformity of PIT values in all three cases; however, this is the consequence of numerical problems resulted by the very large sample size (14566 valid forecast cases), see e.g. \citet{bl15}. The mean $p$-values of 1000 samples from PIT of sizes 1000 each given given in Table \ref{KStab}, nicely reflect to the shapes of the PIT histograms of Figure \ref{pit}.

\section{Conclusions}
\label{sec5}
Various statistical post-processing methods are applied to 3 hourly ensemble forecasts of surface temperature for Santiago de Chile produced by separate runs of the WRF model with 9 different configurations. The one day ahead predictions for different forecast hours are treated separately. The predictive performance of EMOS and BMA models with regional parameter estimation using a 20 day training period is investigated. This optimal length of the training period is a result of a detailed data analysis. Besides the regional models, the forecast skill of an EMOS approach with parameter estimation based on clustering the observation stations is also tested. Compared to the raw ensemble, all post-processing methods for all forecast hours result in a significant decrease in CRPS values of probabilistic forecasts and MAE and RMSE values of point forecasts. In addition, they also yield a substantial improvement in calibration. From the competing calibration approaches the clustering based EMOS produces the smallest score values for all forecast hours; however, the differences in overall mean CRPS and MAE of post-processing models are not significant. We therefore conclude that post-processing of WRF ensemble forecasts for temperature significantly improves the calibration of probabilistic and the accuracy of point forecasts. However, there is still room for improvement, by considering models including spatial dependence via state of the art approaches such as Markovian EMOS \citep{mtlg} or ensemble copula coupling \citep{schefzik16a,schefzik16b}.

\bigskip
\noindent
{\bf Acknowledgments.} \
Mailiu Di\'az is grateful for the support of the National Commission for Scientific and Technological Research (CONICYT) of Chile under Grant No. 21150227.
S\'andor Baran acknowledges the support of the J\'anos Bolyai Research Scholarship of the Hungarian Academy of Sciences and the National Research, Development and Innovation Office under Grant No. NN125679.
Orietta Nicolis and Julio C\'esar Mar\'\i n are partially supported by the Interdisciplinary Center of Atmospheric and Astro-Statistical Studies.
Powered@NLHPC: This research was partially supported by the supercomputing infrastructure of the National Laboratory for High Performing Computer (NLHPC) (ECM-02).


\begin{thebibliography}{99}
\bibitem[Baran, 2014]{bar} Baran, S. (2014) Probabilistic wind speed
  forecasting using Bayesian model averaging with truncated normal components.
  {\em Computational Statistics and Data Analysis\/} {\bf 75}, 227--238.

\bibitem[Baran {\em et al.\/}, 2014]{bhn14} Baran, S., Hor\'anyi, A. and
  Nemoda, D. (2014) Probabilistic temperature forecasting with statistical
  calibration in Hungary. {\em  Meteorology and Atmospheric Physics\/}
  {\bf 124}, 129--142.

\bibitem[Baran and Lerch, 2015]{bl15} Baran, S. and  Lerch, S. (2015)
  Log-normal distribution based EMOS models for probabilistic wind speed
  forecasting. {\em Quarterly Journal of the Royal Meteorological Society\/}
  {\bf 141}, 2289--2299.

\bibitem[Baran and Lerch, 2018]{bl18} Baran, S. and  Lerch, S. (2018)
  Combining predictive distributions for statistical post-processing of
  ensemble forecasts. {\em International Journal of Forecasting\/} {\bf 34},
  477--496.

\bibitem[Baran and Nemoda, 2016]{bn} Baran, S. and Nemoda, D. (2016)
  Censored and shifted gamma distribution based EMOS model for probabilistic
  quantitative precipitation forecasting. {\em Environmetrics\/} {\bf 27},
  280--292.
  
\bibitem[Buizza  {\em et al.\/}, 2005]{bhtp} Buizza, R., Houtekamer, P. L.,
  Toth, Z., Pellerin, G., Wei, M. and Zhu, Y. (2005) A comparison of the ECMWF,
  MSC, and NCEP global ensemble prediction systems.  {\em Monthly Weather
    Review\/} {\bf 133}, 1076--1097.

\bibitem[Burger {\em et al.\/}, 2018]{Burger2018} Burger, F., Brock, B. and
  Montecinos, A. (2018) Seasonal and elevational contrasts in temperature
  trends in Central Chile between 1979 and 2015. {\em Global and Planetary
    Change\/} {\bf 162}, 136--147.

\bibitem[Chen and Dudhia, 2001]{Chen2001} Chen, F. and Dudhia, J. (2001)
  Coupling an advanced land surface-hydrology model with the Penn State-NCAR
  MM5 Modeling System. Part I: Model implementation and sensitivity. {\em
    Monthly Weather  Review\/} {\bf 129}, 569--585.

\bibitem[Cort\'es and Cur\'e, 2011]{Cortes2011} Cort\'es, L. and Cur\'e, M.
  (2011) Validation of the vertical profiles of three meteorological models
  using radiosondes from Antofagasta, Paranal and Llano de Chajnantor. In
  Cur\'e, M., Ot\'arola, A., Mar\'\i n, J. and Sarazin, M. (eds.) {\em
    Astronomical Site Testing Data in Chile.\/} Revista Mexicana de
  Astronom\'\i a y Astrof\'\i sica (Serie de Conferencias) {\bf 41}, pp. 64--67.

\bibitem[Diebold and Mariano, 1995]{dm95} Diebold F. X. and Mariano, R. S.
 (1995) Comparing predictive accuracy. 
  {\em Journal of Business \& Economic Statistics\/} {\bf 13}, 253--263.
  
\bibitem[Dudhia, 1996]{Dudhia1996} Dudhia, J. (1996) A multi-layer soil
  temperature model for MM5. {\em Sixth PSU/NCAR Mesoscale Model Users'
    Workshop\/}, Boulder, 22--24 July 1996, pp. 49--50.

\bibitem[Fraley {\em et al.\/}, 2011]{frgsb} Fraley, C., Raftery,
  A. E., Gneiting, T., Sloughter, J. M. and Berrocal, V. J. (2011)
  Probabilistic weather forecasting in R.  {\em The R Journal\/} {\bf
    3}, 55--63. 

\bibitem[Gebhardt {\em et al.\/}, 2011]{gtpb} Gebhardt, C., Theis, S. E.,
  Paulat, M. and Ben Bouall\`egue, Z. (2011) Uncertainties in COSMO-DE
  precipitation forecasts introduced by model perturbations and variation of
  lateral boundaries. {\em Atmospheric Research\/} {\bf 100}, 168--177.

\bibitem[Gneiting, 2011]{gneiting11} Gneiting, T. (2011) Making and
  evaluating point forecasts.  {\em Journal of the American Statistical
    Association\/} {\bf 106}, 746--762.
  
\bibitem[Gneiting {\em et al.\/}, 2007]{gbr} Gneiting, T.,
  Balabdaoui, F. and Raftery, A. E. (2007) Probabilistic forecasts,
  calibration and sharpness. {\em Journal of the Royal Statistical Society:
    Series B\/} {\bf 69}, 243--268.

 \bibitem[Gneiting and Raftery, 2007]{grjasa} Gneiting, T. and Raftery,
  A. E. (2007) Strictly proper scoring rules, prediction and
  estimation. {\em Journal of the American Statistical Association\/}
  {\bf 102}, 359--378.

\bibitem[Gneiting {\em et al.\/}, 2005]{grwg} Gneiting, T., Raftery, A. E.,
  Westveld, A. H. and Goldman, T. (2005) Calibrated probabilistic forecasting
  using ensemble model output statistics and minimum CRPS estimation.
  {\em Monthly Weather Review\/} {\bf 133}, 1098--1118.
  
\bibitem[Gonz\'alez and Garreaud, 2017]{Gonzalez2017} Gonz{\'a}lez, S. and
  Garreaud, R. (2017) Spatial variability of near-surface temperature over the
  coastal mountains in southern Chile (38{$^{\circ}$S}) {\em  Meteorology and
    Atmospheric Physics\/}, \ https://doi.org/10.1007/s00703-017-0555-4.

\bibitem[Hemri {\em et al.\/}, 2014]{hemri14} Hemri, S., Scheuerer, M., 
  Pappenberger, F., Bogner, K. and Haiden, T. (2014) Trends in the predictive 
  performance of raw ensemble weather forecasts. {\em Geophysical Research
    Letters\/}  {\bf 41}, 9197--9205.
  
\bibitem[Hong {\em et al.\/}, 2004]{Hong2004} Hong, S.-Y., Dudhia, J. and
  Chen, S.-H. (2004) A revised approach to ice microphysical processes for the
  bulk parameterization of clouds and precipitation. {\em Monthly Weather
    Review\/} {\bf 132}, 103--120.
  
\bibitem[Hong {\em et al.\/}, 2006]{Hong2006} Hong, S.-Y., Noh, Y. and Dudhia,
  J. (2006) A new vertical diffusion package with an explicit treatment of
  entrainment processes. {\em Monthly Weather Review\/} {\bf 134}, 2318--2341.

\bibitem[Iacono {\em et al.\/}, 2005]{Iacono2008} Iacono, M. J., Delamere
  J. S., Mlawer, E. J., Shephard, M. W., Clough, S. A. and Collins, W. D. (2005)
  Radiative forcing by long-lived greenhouse gases: Calculations with the AER
  radiative transfer models. {\em Journal of Geophysical Research:
    Atmospheres\/} {\bf 113}, D13103.

\bibitem[Janji\'{c}, 1994]{Janjic1994} Janji\'{c}, Z. I. (1994) The
  Step-Mountain Eta Coordinate Model: further developments of the convection,
  viscous sublayer, and turbulence closure schemes. {\em Monthly Weather
    Review\/} {\bf 122}, 927--945.

\bibitem[Kain, 2004]{Kain2004} Kain, J. S. (2004) The Kain-Fritsch convective
  parameterization: an update. {\em Journal of Applied Meteorology\/} {\bf 43},
  170--181.

\bibitem[Leith, 1974]{leith} Leith, C. E. (1974) Theoretical skill of
  Monte-Carlo forecasts.  {\em Monthly Weather Review\/} {\bf 102}, 409--418.

\bibitem[Lerch and Baran, 2017]{lb17} Lerch, S. and Baran, S. (2017) 
  Similarity-based semi-local estimation of EMOS models. {\em Journal
    of the Royal Statistical Society: Series C\/} {\bf 66}, 29--51.

\bibitem[Leutbecher and Palmer, 2008]{lp08} Leutbecher, M. and Palmer,
  T. N. (2008) Ensemble forecasting. {\em Journal of Computational Physics\/}
  {\bf 227}, 3515--3539.

\bibitem[Mar\'\i n {\em et al.\/}, 2013]{mpmtc13} Mar\'\i n, J. C., Pozo, D.,
  Mlawer, E., Turner, D. and Cur\'e, M. (2013) Dynamics of local circulations
  in mountainous terrain during the RHUBC-II project. {\em Monthly Weather
    Review\/} {\bf 141}, 3641--3656.

\bibitem[Massey {\em et al.\/}, 2016]{mskc16} Massey, J. D., Steenburgh,  W.
  J., Knievel, J. C. and Cheng, W. Y. Y. (2016) Regional soil moisture biases
  and their influence on WRF Model temperature forecasts over the Intermountain
  West. {\em Weather and Forecasting\/} {\bf 31}, 197--216,

\bibitem[McLachlan and  Krishnan, 1997]{mclk} McLachlan, G. J. and  Krishnan,
  T. (1997) {\em The EM Algorithm and Extensions.\/} Wiley, New York. 

\bibitem[Molteni {\em et al.\/}, 1996]{ecmwfens} Molteni, F., Buizza, R.,
  Palmer, T. N. and Petroliagis, T. (1996) The ECMWF ensemble prediction
  system: Methodology and validation. {\em Quarterly Journal of the Royal
    Meteorological Society\/} {\bf 122}, 73--119.

\bibitem[M\"oller {\em et al.\/}, 2015]{mtlg} M\"oller, A., Thorarinsdottir,
  T. L., Lenkoski, A. and Gneiting, T. (2015) Spatially adaptive, Bayesian
  estimation for probabilistic temperature forecasts. arXiv:1507.05066. 

\bibitem[Nakanishi and Niino, 2006]{Nakanishi2006} Nakanishi, M. and Niino, H.
  (2006) An improved Mellor--Yamada Level-3 Model: its numerical stability and
  application to a regional prediction of advection fog. {\em Boundary-Layer
    Meteorology\/} {\bf 119}, 397--407.

\bibitem[Pinson and Hagedorn, 2012]{pinhag} Pinson, P. and Hagedorn,
  R. (2012) Verification of the ECMWF ensemble forecasts of wind speed
  against analyses and observations. {\em Meteorological Applications\/}
  {\bf 19}, 484--500. 

\bibitem[Pleim and Xiu, 2003]{Pleim2003} Pleim, J. E. and Xiu, A. (2003)
  Development of a land surface model. Part II: Data assimilation. {\em
    Journal of Applied Meteorology\/} {\bf 42}, 1811-1822.
  
\bibitem[Powers {\em et al.\/}, 2017]{wrfmod} Powers, J. G., Klemp, J. G.,
  Skamarock, W. S., Davis, C. A., Dudhia, J., Gill, D. O., Coen, J. L.,
  Gochis, D. J., Ahmadov, R., Peckham, S. E., Grell, G. A., Michalakes, J.,
  Trahan, S., Benjamin, S. G., Alexander, C. R., Dimego, G. J., Wang, W.,
  Schwartz, C. S., Romine, G. S., Liu, Z., Snyder, C., Chen, F., Barlage, M.
  J., Yu, W. and Duda, M. G. (2017) The Weather Research and Forecasting Model:
  overview, system efforts, and future directions. {\em Bulletin of the
    American Meteorological Society\/} {\bf 98}, 1717--1737.

\bibitem[Pozo {\em et al.\/}, 2016]{Pozo2016} Pozo, D., Mar\'{i}n, J. C.,
  Illanes, L., Cur\'{e}, M. and Rabanus, D. (2016) Validation of {WRF}
  forecasts for the Chajnantor region. {\em Monthly Notices of the Royal
    Astronomical Society\/} {\bf 459}, 419--426.

\bibitem[Raftery {\em et al.\/}, 2005]{rgbp} Raftery, A. E., Gneiting, T.,
  Balabdaoui, F. and Polakowski, M. (2005) Using Bayesian model averaging to
  calibrate forecast ensembles. {\em Monthly Weather Review\/} {\bf 133},
  1155--1174.

\bibitem[Ruiz {\em et al.\/}, 2010]{rsnp10}  Ruiz, J. J., Saulo, C. and
  Nogues-Paegle, J. (2010) WRF model sensitivity to choice of parameterization
  over South America: Validation against surface variables. {\em Monthly
    Weather Review\/} {\bf 138}, 3342--3355.

\bibitem[Saide {\em et al.\/}, 2011]{Saide2011} Saide, P. E., Carmichael, G.
  R., Spak, S. N., Gallardo, L., Osses, A. E., Mena-Carrasco, M. A. and
  Pagowski, M. (2011) Forecasting urban PM10 and PM2.5 pollution episodes in
  very stable nocturnal conditions and complex terrain using WRF–Chem CO tracer
  model. {\em Atmospheric Environment\/} {\bf 45}, 2769--2780.

\bibitem[Schefzik, 2016a]{schefzik16a} Schefzik, R. (2016a) A
  similarity-based implementation of the Schaake shuffle. {\em Monthly Weather
    Review\/} {\bf 144}, 1909--1921.
  
\bibitem[Schefzik, 2016b]{schefzik16b} Schefzik, R. (2016b) Combining parametric
  low-dimensional ensemble postprocessing with reordering methods. {\em
    Quarterly Journal of the Royal Meteorological Society\/} {\bf 142},
  2463--2477.
  
\bibitem[Scheuerer, 2014]{sch} Scheuerer, M. (2014) Probabilistic quantitative
  precipitation forecasting using ensemble model output statistics. {\em
    Quarterly Journal of the Royal Meteorological Society\/} {\bf 140},
  1086--1096.

\bibitem[Scheuerer and Hamill, 2015]{schham} Scheuerer, M. and Hamill, T. M.
  (2015) Statistical post-processing of ensemble precipitation forecasts by
  fitting censored, shifted gamma distributions.  {\em Monthly Weather
    Review\/}  {\bf 143}, 4578--4596.

\bibitem[Skamarock {\em et al.\/}, 2008]{Skamarock2008} Skamarock, W. C.,
  Klemp, J. B., Dudhia, J., Gill, D. O., Barker, D. M., Duda, M., Huang, X. Y.,
  Wang, W. and Powers, J .G. (2008). A description of the Advanced Research WRF
  Version 3. NCAR Technical Note, NCAR/TN–475+ST.

\bibitem[Sloughter {\em et al.\/}, 2010]{sgr10} Sloughter, J. M., Gneiting, T.
  and Raftery, A. E. (2010) Probabilistic wind speed forecasting using
  ensembles and Bayesian model averaging. {\em Journal of the American
    Statistical Association\/} {\bf 105}, 25--37.

\bibitem[Sloughter {\em et al.\/}, 2007]{srgf} Sloughter, J. M., Raftery, A.
  E., Gneiting, T. and Fraley, C. (2007) Probabilistic quantitative
  precipitation forecasting using Bayesian model averaging. {\em Monthly
    Weather Review\/} {\bf 135}, 3209--3220.

\bibitem[Thorarinsdottir and Gneiting, 2010]{tg}  Thorarinsdottir, T. L. and
  Gneiting, T. (2010) Probabilistic forecasts of wind speed: ensemble model
  output statistics by using heteroscedastic censored regression. {\em Journal
    of the Royal Statistical Society: Series A\/} {\bf 173},  371--388. 

\bibitem[Wannitsem {\em et al.\/}, 2018]{ppb}  Wannitsem, S., Wilks, D. S. and
  Messner, J. W. (eds.) (2018) {\em Statistical Postprocessing of Ensemble
    Forecasts\/}, Elsevier, Amsterdam.
  
\bibitem[Williams {\em et al.\/}, 2014]{wfk} Williams, R. M., Ferro,
  C. A. T. and  Kwasniok F (2014) A comparison of ensemble post-processing
  methods for extreme events. {\em  Quarterly Journal of the Royal
    Meteorological Society\/} {\bf 140}, 1112--1120.

\bibitem[Wilks, 2011]{wilks} Wilks, D. S. (2011) 
   {\em Statistical Methods in the Atmospheric Sciences\/} (3rd ed.).
   Elsevier,  Amsterdam.

\bibitem[Zhang {\em et al.\/}, 2018]{zhang18} Zhang, Y. L., Li, X., Cheng, G.
  D., Jin, H. J., Yang, D. W., Flerchinger, G. N., Chang, X. L., Wang, X. and
  Liang, J. (2018) Influences of topographic shadows on the thermal and
  hydrological processes in a cold region mountainous watershed in northwest
  China. {\em Journal of Advances in Modeling Earth Systems\/} {\bf 10},
  1439--1457.

\end{thebibliography}

\end{document}